\documentclass[english]{article}
\pdfoutput=1
\usepackage[utf8]{inputenc}
\usepackage[T1]{fontenc}
\let\tldocfrench=1 
\usepackage{csquotes}
\usepackage{babel}
\usepackage{xspace}
\usepackage{amsmath}
\usepackage{amsfonts}
\usepackage{amssymb}
\usepackage{tabularx}
\usepackage{array}
\usepackage{cases}
\usepackage{graphicx}
\usepackage{lmodern}
\usepackage{makeidx}
\usepackage{xkeyval}

\usepackage{tex-live}
\ifpdf
\usepackage{hyperref}
\usepackage[all]{hypcap}
\fi
\begin{document}

\title{%
  \textbf{Detailed Electron Energy Loss Spectroscopy (EELS) Microanalysis of Data Collected Under Semi-Angle Less Than Both Plasmon Cutoff Angle and Incident Beam Convergence Semi-Angle.}%
}
\hypersetup{pdfauthor={some author},pdftitle={eye-catching title}}
\author{
  Noureddine HADJI %
}
\date{September 2020}
\maketitle
\pagenumbering{arabic}
\begin{center} \end{center}
\noindent D\'{e}partement de Physique, Universit\'{e} Badji Mokhtar, Annaba, BP 12 Annaba 23000, Alg\'{e}rie.\\
\textit{Alternative address:} 138, Villa 60, Cit\'{e} des Jardins, EL HADJAR, Annaba 23200, Algeria.\\
{\it Email:} noureddine.hadji@univ-annaba.dz \\
\includegraphics[scale=0.28]{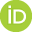}  \href{https://orcid.org/0000-0001-5876-8002}{https://orcid.org/0000-0001-5876-8002}
\\
\pagenumbering{gobble} 
\pagenumbering{arabic}
\\
\noindent In previous work a different and powerful, analytical, technique was used to get data, such as the absolute atom concentration (AAC), specimen thickness etc., from public domain boron nitride EELS spectrum collected under a collection semi-angle, $\beta$, less than the plasmon cutoff angle, $\theta_{c}$, but large relative to incident beam convergence, $\alpha$. Here, seeking for some completeness, another, numerical, technique usable together with, also, $\beta < \theta_{c}$ is described in \emph{minute detail} and applied to data obtained with $\beta/\alpha < 2$, so necessitating incident beam convergence-related corrections. A lot of experimental physical parameters all fully relevant to one another are produced from a single EELS spectrum. Of public domain silicon nitride, Si$_{3}$N$_{4}$, EELS spectrum used. Comparison between results producible by the two $\beta<\theta_{c}$-related techniques made. Results range from parameters such as AAC, density, plasmon critical vector, plasmon dispersion coefficient, Fermi energy to specimen thickness. Results were obtained using version 5 of eelsMicr program and compared with existing results obtained using non EELS techniques.  \\
%
\begin{multicols}{2}
  \tableofcontents
  \listoffigures
\listoftables
\end{multicols}
\section{Introduction}
In (most) real specimens, in practical EELS, polyscattering not only occurs but is also accompanied with plural  scattering. The combination of these two gives rise to components in EELS spectra made up of electrons that have experienced more than one scattering event, some of which (components) are made up of electrons that have experienced events related to different types of scattering processes (e.g. one plasmon scattering event and one inner shell, e.g. K-shell, scattering event) thus leading to rather complex spectra to which one cannot apply the `single scattering' formulae  usually used in EELS microanalysis without first removing the unwanted effects of the combination of plural scattering and polyscattering. The problem relative to these effects was fully dealt with in a work reported elsewhere, see \cite{Hadji-2018a}, and is only considered here with the aim of giving a sort of summary, of the reported work, which is easier to use than the mother text when the work under consideration is only concerned with EELS and physical data extractions from experimental spectra. Effectively, a detailed account of a study, the work reported, carried out with the goal of getting to an EELS microanalysis usable also for quantitative chemical characterization of materials through EELS investigation of thin specimens with much larger thicknesses was presented and validated in \cite{Hadji-2018a}. Part of the validation was through the use of a version of Technique 1, the first of the two analytical techniques described in \cite{Hadji-2002} and applied in \cite{Hadji-2016} and \cite{Hadji-2015}. The main aim of the present work is to detail and apply a slightly different analytical procedure, a procedure connected with another possible version of Technique 1 in which the solution of the relevant system of equations is numerical. This second version requires the electron effective mass is known and allows producing, through calculation, an experimental value for the plasmon energy and comparing it with the value measurable directly from the considered EELS spectrum. This second version may also prove to be useful in studies, using EELS, in which the determination of the electron effective mass is through means other than EELS. Effcetively, the procedure employed in \cite{Hadji-2018a} to get the AAC from an experimental EELS spectrum collected under a collection semi-angle less than the plasmon cutoff angle is that which corresponds to the first, Technique 1, of the two analytical techniques described, in connection with a collection semi-angle larger than the plasmon cutoff angle, in \cite{Hadji-2002} and applied in \cite{Hadji-2016} and \cite{Hadji-2015}. Therefore, a technique for the determination of the absolute atom concentration which (i) considers all chemical elements present within the specimen under study are easily detectable, (ii) uses a measure of $E_{p}$, the plasmon energy or first plasmon loss peak position, but does not rely on the formula for this $E_{p}$ and (iii)  does not require the knowledge of a value for the electron effective mass was used in \cite{Hadji-2018a}. A consequence of this is that the electron effective mass, $m$, could be determined through an equation formed using the usual formula for the plasmon energy and the value for this energy measurable from the experimental EELS spectrum. Here, the technique considered for the determination of the absolute atom concentration is that which is usable together with EELS spectra collected, also, under collection semi-angles less than the plasmon cutoff angle and which connects also with Technique 1 but which (j) uses a formula for the plasmon energy instead of an experimental value, (jj) assumes the electron effective mass, $m$, is available and (jjj) obtains a value for the plasmon energy through (\ref{02Two}), see below, using the available $m$ and the determined electron concentration, $n_{e}$. The determination of $n_{e}$ is through making use of the below (\ref{10Ten}), obtained on combining the, see below also, (\ref{07Seven}), (\ref{02Two}), (\ref{06Six}), (\ref{04Four}) and (\ref{09Nine}), in which an experimental value for the plasmon scattering cross section per atom of species $k$, $\sigma_{p}^{k} (\beta)$, and a value for the electron effective mass, $m$, are inserted before solving numerically the resulting equation to get an experimental value for $n_{e}$ and then getting through (\ref{02Two}) another, different, experimental value for $E_{p}$. Thus in this second procedure the same formulas as those used in \cite{Hadji-2018a} are combined to form a different equation whose solution is obtained numerically, not analytically.
Finally, the experimental EELS data considered here fall within the category of the EELS data for which incident beam convergence-related corrections are necessary before use together with the usual equations determined for a parallel illumination, e.g. those obtained and used in \cite{Hadji-2018a}.
\section{Other version of Technique 1, a description}
\label{2Techniq1}
When the collection semi-angle is less than the plasmon cutoff angle another version of Technique 1 is possible. This version is described in this section.\\
The differences between the way of applying Technique 1 together with EELS data collected with a collection semi-angle, $\beta$, larger than the plasmon cutoff angle, $\theta_{c}$, and the way of applying it together with EELS data collected under a $\beta$ less than $\theta_{c}$ come from the replacement in the former case of $\beta$ by $\theta_{c}$ in the formula relative to the plasmon mean free path given by, e.g. see \cite{Egerton-1996}: 
\begin{equation}
\lambda_{p}\left( \beta\right)=\frac{2a_{0}m_{0}v^{2}}{E_{p}\ln \left( 1+\frac{\beta^{2}}{\left( \theta_{E}\right) ^{2}}\right) }=\frac{2a_{0}}{ \gamma\theta_{E}\ln \left( 1+\frac{\beta^{2}}{\left( \theta_{E}\right) ^{2}}\right) },  \label{01One} 
\end{equation}
plus using a formula for $\theta_{E}$, the characteristic angle, and using in the latter case an experimental value for  $\beta$ together with either, so also, a formula or an experimental value for $\theta_{E}$. The sub case, or analytical version of Technique 1, using an experimental value for $\theta_{E}$ is  the sub case, or analytical version of Technique 1, utilized in \cite{Hadji-2018a}. In both of the former case and the remaining sub case of the latter case, or numerical version of Technique 1 - this is the version described in detail in the present work-, the formula used for $\theta_{E}$ is that given by (\ref{06Six}) below in which the electron concentration $n_{e}$, given by (\ref{04Four}), relates to, but does not necessarily represent, the AACs of the various chemical species present within the irradiated specimen volume. This is so, see \cite{Hadji-2016}, for the reason that $n_{e}$ can represent the concentration of the valence electrons associated with the atoms constituting the specimen matrix and cannot represent the valence electrons of the atoms not among those constituting the matrix, i.e. of the atoms not bonded to the matrix, e.g. those, if any, making up inclusions present under molecular forms within the irradiated specimen volume and, thus, those not providing the collective of electrons whose concentration is $n_{e}$ with their outer electrons, i.e. with their valence electrons.
\{In (\ref{01One}) $a_{0}$ is the Bohr radius, $m_{0}$ the electron rest mass, $v$ the incident electron velocity, $E_{p}$ is the plasmon loss energy and is given by:
\begin{equation}
E_{p}=\hbar \sqrt {\left( \frac{n_{e}e^{2}}{m\varepsilon_{0}}\right)}, \label{02Two}
\end{equation}
[in (\ref{02Two}) $\hbar=h/2\pi$  where $h$ is Planck's constant, $n_{e}$, the \index{electron concentration}electron concentration, is given by
\begin{equation}
n_{e}=\sum_{q=1}n_{q}p_{q}, \label{03Three}
\end{equation}
(where $n_{q}$ and $p_{q}$ are, respectively, the AAC and the number of the valence electrons of the $q$th atom species), $e$ the electron charge, $m$ is the electron effective mass, and $\varepsilon_{0}$ the permittivity of space; this $n_{e}$ can also be expressed in terms of the content ratios relative to the chemical species constituting the material's matrix of the specimen region under consideration if the material under consideration is not monatomic:
\begin{equation}
n_{e}=n_{k}\sum_{q=1} \frac{n_{q}}{n_{k}}  p_{q} = n_{k} \left(p_{k}+\sum_{q=1,q\neq k} \frac{n_{q}}{n_{k}}  p_{q}\right),  \label{04Four}    
\end{equation}
where $n_{q}/n_{k}$ is the content ratio of the $q$th chemical species relative to the chemical species $k$; this ratio is also given by, see \cite{Hadji-2002},:
\begin{equation}
\frac{n_{q}}{n_{k}} =\frac{\sigma_{p}^{k}}{\sigma_{p}^{q}},    \label{05Five}
\end{equation}
where $\sigma_{p}^{k}$ and $\sigma_{p}^{q}$ are the plasmon scattering cross sections (PSCA)s, respectively, for the $k$ and the $q$ atom species and can be obtained from (\ref{07Seven}) below;],
\begin{equation}
\theta_{E}=\frac{E_{p}}{\gamma m_{0} v^{2}}=\frac{\hbar }{\gamma m_{0} v^{2}} \sqrt{\frac{n_{e} e^{2}}{m\varepsilon_{0}}}, \label{06Six}
\end{equation}
where $\gamma = 1/\sqrt{1-v^{2}/c^{2}} $, $c$ being the speed of light in free space.\}
\noindent Therefore, in the sub case described in the work considered here, i.e. the numerical version of Technique 1 when $\beta < \theta_{c}$, the PSCA for the species $k$, $\sigma_{p}^{k} (\beta)$, is given by
\begin{equation}
\sigma_{p}^{k} (\beta)=\frac{\gamma \theta_{E} }{2a_{0}n_{k}}\ln \left(1+\frac{\beta^{2}}{\theta_{E}^{2}} \right),  \label{07Seven} 
\end{equation}
since $\sigma_{p}^{k}$ and $\lambda_{p}$ are correlated by :
\begin{equation}
\sigma_{p}^{k}=\frac{1}{n_{k}\lambda_{p}}.  \label{08Eight}
\end{equation}
Consequently, the knowledge of the various \emph{right} $\sigma_{p}^{k} (\beta)$ values relative to the various $k$ chemical species allows 
\begin{description}
\item[(i)] expressing $n_{e}$, the electron concentration, given by (\ref{04Four}) in terms of a unique unknown AAC, e.g. $n_{k}$, and therefore $n_{k}$ in terms of $n_{e}$ as: 
\begin{equation}
n_{k}= \frac{n_{e}}{\left( p_{k}+\sum_{q=1,q\neq k}\frac{n_{q}}{n_{k}}  p_{q} \right)},  \label{09Nine}  
\end{equation}
and thereafter
\item[(ii)] obtaining through (\ref{07Seven}), (\ref{06Six}), (\ref{04Four}) and (\ref{09Nine}) an expression of $\sigma_{p}^{k} (\beta)$, the PSCA for the $k$ species as:
\begin{equation}
\sigma_{p}^{k} (\beta)=\frac{\hbar e \left( p_{k}+\sum_{q=1,q\neq k}\left[\frac{n_{q}}{n_{k}} \right] p_{q} \right) }{4a_{0} T(m\varepsilon_{0})^{1/2}n_{e}^{1/2}} \ln\left(1+\frac{\beta^{2}}{n_{e}\frac{\hbar^{2} e^{2}}{\left( 2 \gamma T\right)^{2}  m\varepsilon_{0}}}\right), T=\frac{m_{0} v^{2}}{2} \label{10Ten}
\end{equation}
which is also a function of the electron concentration and the electron effective mass. This, in turn, leads to a non-linear equation in $n_{e}$ with just one unknown if a value for $\sigma_{p}^{k} (\beta)$ is available together with a value for $m$, obtained, for instance, experimentally through means other than EELS. Solving, numerically, this equation in $n_{e}$, is therefore equivalent to the experimental determination of the electron concentration and, thus, to the determination, through (\ref{09Nine}), of $n_{k}$, the AAC relative to the $k$th chemical species. The use, thereafter, of (\ref{05Five}) will allow producing the AAC of the remaining chemical elements, since the various $\sigma_{p}^{k} (\beta)$ can be made available experimentally. After that, obtaining a number of physical parameters, e.g. specimen thickness, plasmon mean free path and so forth, becomes straightforward.
\end{description}
Thus obtaining the right values for $\sigma_{p}^{k} (\beta)$ and for the content ratios is required to form the non-linear equation from which to deduce the right electron concentration. The experimental $\sigma_{p}^{k} (\beta)$ can be obtained, see \cite{Hadji-2002, Hadji-2016}, from measures of the two EELS parameters $n_{k}^{a}$, the NAA of $k$ chemical species, and $t/\lambda_{p}$, the thickness to plasmon mean free path ratio through: 
\begin{equation}
\sigma_{p}^{k} (\beta)=\frac{t/(\lambda_{p}(\beta))}{n_{k}^{a}}\label{11Eleven}
\end{equation}
in which the right values for $t/\lambda_{p}(\beta) $ and $n_{k}^{a}$, see \cite{Hadji-2018a}, are inserted, these right values being values obtained either directly or indirectly under experimental conditions, on the one hand, similar to those assumed\footnote{The experimental conditions assumed when the said formulas were obtained were, in reality, tacitly assumed and consisted in considering that the inelastic scattering were through just one inelastic scattering process at a time. Examples: a) (\ref{01One}) and (\ref{07Seven}) were obtained assuming the inelastic scattering were through the plasmon scattering process only, and thus considered that the other possible inelastic scattering processes, e.g. the inner-shell-related inelastic scattering processes, were not active, were not on, and b) $n_{k}^{a}$, the NAA for the atom species $k$, need be obtained under conditions as similar as possible to those which need be considered in connection with the relevant experimental intensity required to get the value for this $n_{k}^{a}$ which is necessary for validly using (\ref{11Eleven}), that is conditions in which the inelastic scattering would be through the relevant inner-shell inelastic scattering process only, with the other possible inelastic scattering processes inactive, switched off.} during the theoretical processes of derivation of equations from (\ref{01One}) to (\ref{10Ten}) and, on the other hand, to be satisfied in connection with the calculation of the value of $n_{k}^{a}$; ($\alpha$) a direct determination of the right values would involve acquisitions of spectra each of which would be relative to inelastic scattering through no more than one inelastic scattering process, a thing not easy to achieve indeed, and ($\beta$) an indirect determination of the right values can be embodied by their extractions, see \cite{Hadji-2018a} and below, from EELS spectra obtained under the real experimental conditions prevailing when real EELS experiments are carried out, which is much easier indeed. Thus $t/\lambda_{p}(\beta) $ is the ``true'' experimental $t/\lambda_{p}(\beta) $, that is the experimental value for $t/\lambda_{p}(\beta)$ which would be observed if the scattering were through the plasmon inelastic scattering process only and the experimental $n_{k}^{a} (=t n_{k},)$ is the also ``true'' $n_{k}^{a}$ and, thus, is that which would be obtained if the inelastic scattering were through just the inner-shell inelastic scattering process associated with the generation of the inner-shell edge the inelastically scattered intensity used to get this $n_{k}^{a}$ is from.  This is so for the reason that (\ref{07Seven}) and, therefore, (\ref{01One}) are for a system of electrons, the valence electrons whose concentration is $n_{e}$, acting collectively through their 'inelastic scattering capacity' on an incident monoenergetic electron beam in absence of interaction of the beam with any other possible inelastic scattering process-related entity; therefore, $\lambda_{p}(\beta)$ represents the mean distance between two successive inelastic scattering events experienced by an electron incident onto the system of electrons on interacting with it, i.e. the mean free path known as the plasmon mean free path,\footnote{Thus, this is basically an electron inelastic scattering mean free path which could be named ``electron (plasmon) inelastic scattering mean free path'' but which is totally different from other inelastic scattering mean free paths such as, for example, that relative to electrons inelastically scattered in connection with transport phenomena.} in absence of inelastic scattering through other, possible, inelastic scattering processes relevant to the specimen under study. \\
\noindent The true experimental values for $t/\lambda_{p}(\beta) $  and $n_{k}^{a}$ can be obtained, see \cite{Hadji-2018a} and below, by submitting the corresponding, ``starred'', values extractible, using the ``successive approximation methos (SAM)'', from a real experimental EELS spectrum to corrections for the unwanted effects of the combination of the plural\footnote{Plural scattering is the scattering which involves a particle in at least two scattering events, for instance on traversing a specimen.} and poly\footnote{By definition we consider ‘polyscattering’ or `poly scattering' to designate the scattering occurring in situations where the interactions between incident electron beams and material targets involve more than one type of scattering processes. For instance, the processes through which plasmon scattering and the scattering by inner shells occur are two types of such scattering processes.} scatterings phenomena. The said corresponding starred values are, thus, relative to $t/\lambda_{p}(\beta) $ and $n_{k}^{a}$ but are obtainable from the single inelastic scattering intensities of the appropriate EELS spectrum components of the real EELS spectrum. That is from the spectrum components which have been determined by the contribution of all of the inelastic scattering processes relevant to the material under study and not just by the spectrum components each of which is relative to one of the different EELS spectra which would be obtained experimentally in absence of polyscattering, The true data are thus the results of the application of the full improved EELS microanalysis method (IEELSM), see \cite{Hadji-2018a}, and their use is for producing valuable physical parameters, for instance starting with the electron concentration, which are proper to the material they relate to.
\section{Getting the electron concentration}
\label{GettElecConcent}
Two categories of EELS data are possible. One of these includes EELS data that do not require corrections for incident beam convergence, and thus includes all of the data obtainable under parallel illumination and part of the data obtainable using convergent incident beams, those with collection semi-angle, $\beta$, to the incident beam convergence semi-angle,$\alpha$, ratios larger than 2 (so $\beta/\alpha>2$), see \cite{Egerton-1996}. The second category encompasses all of the EELS data with $\beta/\alpha\leq2$, this knowing that models for performing corrections in connection with the incident beam convergence effects exist, see \cite{Egerton-1996}-\cite{IAKOUBOVSKII-et-al-2008}. However, the expressions by \cite{Scheinfein-Isaacson-1984, Craven-et-al-1981} give corrected inelastically scattered intensities collected with $\beta/\alpha\leq2$ which can immediately be employed to get the proper EELS quantities required for the proper use of SAM to extract the various starred EELS parameters from the real spectrum, obtained with $\beta/\alpha\leq2$, under study; parameters that must, therefore, be free from incident beam convergence effects. Here, the formulas by \cite{Scheinfein-Isaacson-1984}, used through the CONCOR2 program given in \cite{Egerton-1996}, have yielded fully satisfactory results. For data obtained using convergent incident beams not necessitating corrections and for \emph{corrected} data, the essential equation to solve to get the electron concentration in connection with an arbitrary EELS spectrum is formed using (\ref{09Nine}) together with the true experimental PSCAs relative to the different chemical species present within the specimen the arbitrary EELS spectrum originates from, i.e. $\sigma_{p}^{k}(\beta)$, $k = 1$(e.g. = Si), 2(e.g. = N), 3, $\cdots $, alongside with $m$. The determination of these PSCAs requires:
\begin{description}
\item[1)]  \textbf{the extraction of the starred NAAs relative to the various $k$ atom species and the starred thickness to plasmon mean free path ratio}, respectively, $n_{k}^{*,a}$ and $t/\lambda_{p}^{*} $ from the experimental spectrum in consideration. The determination of these $n_{k}^{*,a}$ and $t/\lambda_{p}^{*} $ is achieved here through the use of SAM, see \cite{Hadji-2018a} for the full description of this method, by solving numerically the set of equations, given below, (\ref{12Twelve}), (\ref{14Fourteen}), (\ref{15Fifteen}) and (\ref{16Sixteen}), (these are the ones used in eelsMicr program, see \cite{Hadji-2018b} and \cite{Hadji-2018c} for source code and application associated with this program). These equations relate to:
\begin{itemize}
 \item $n_{k}^{*,a}$, \textbf{\emph{the starred NAAs for the various chemical species $k$}}; this $n_{k}^{*,a}$ can be written as, see \cite{Hadji-2018a}
\begin{equation}
n_{k}^{*,a}= n_{k}^{a,\textrm{NM}}\frac{\sigma_{k,q}\left( \Delta, \beta \right)}{\sigma_{k,q}^{eff}\left( \Delta, \beta \right)} , \label{12Twelve}
\end{equation}
where $n_{k}^{a,\textrm{NM}}$ is the NAA as can be obtained within the framework of the so called normal EELS microanalysis method (NEELSM), e.g. see \cite{Egerton-1996} for this method, and is given by:
\begin{equation}
n_{k}^{a,\textrm{NM}} = \frac{I_{k,q}^{ed}\left( \Delta, \beta \right) }{I_{LL}\left( \Delta, \beta \right)\sigma_{k,q}\left( \Delta, \beta \right)}. \label{13Thirteen}
\end{equation}
{\large \textbf{[}}In this (\ref{13Thirteen}) $I_{k,q}^{ed}\left( \Delta, \beta \right)$ is obtained from the EELS spectrum under consideration and is the sum of the partial intensities $I_{k,q}^{*,j}\left[ \Delta - (j-1)E_{p}, \beta \right]$, $j$ = 1, 2, $\cdots$,  corresponding to the contributions to the $(k,q)$-edge intensity due, on the one hand, to the single inelastic scattering through the $(k,q)$-process, i.e. through the $q$ inelastic scattering process of the atom species $k$ (e.g. $q$ = L-shell-related inelastic scattering process of $k$ = Si, $q$ = K-shell-related inelastic scattering process of $k$ = N etc.), and, on the other hand, to the various plural inelastic scatterings involving one scattering event through the $(k,q)$-process plus one or more, i.e. plus ($j$ - 1), $j$ = 2, 3, $\cdots$, inelastic scattering events through the plasmon process; this $I_{k,q}^{ed}\left( \Delta, \beta \right)$ is measured after estimating and subtracting the relevant underlying background signal; $\Delta$ is the energy interval width, starting at the $(k,q)$-edge threshold energy, over which the $(k,q)$-edge intensity is measured; $\sigma_{k,q}\left( \Delta, \beta \right)$ is the partial cross section of characteristic signal $q$ of atom species $k$ and is a theoretically calculable parameter; $I_{LL}\left( \Delta, \beta \right)$ is the intensity within the low loss energy part of the EELS spectrum extending up to the energy loss $ E = \Delta ${\large \textbf{]}}. Therefore, a way of getting $n_{k}^{*,a}$ consists in using (\ref{12Twelve}). This requires obtaining $I_{k,q}^{ed}\left( \Delta, \beta \right)$ and $I_{LL}\left( \Delta, \beta \right)$, inserting them into (\ref{13Thirteen}) and, finally, inserting the resulting value for $n_{k}^{a,\textrm{NM}}$ into (\ref{12Twelve}) together with $\sigma_{k,q}^{eff}\left( \Delta, \beta\right)$, the $({k,q})$-inner shell related effective scattering cross section, given by, see \cite{Hadji-2018a} for details:
\begin{equation}
\sigma_{k,q}^{eff}\left( \Delta, \beta\right) = \frac{\sum_{j=1}\left[ t/\lambda_{p}^{*}\left( \beta\right)\right]^{j-1}\sigma_{k,q}\left[ \Delta - \left( j-1\right) E_{p}, \beta\right] /\left( j-1\right) !}{\sum_{j=1}\left[ t/\lambda_{p}^{*}\left( \beta\right)\right]^{j-1}/\left( j-1\right) !}. \label{14Fourteen}
\end{equation}
where $\beta$ is the experimental collection semi-angle; $t/\lambda_{p}^{*}\left( \beta\right)(\equiv t/\lambda_{p}^{*})$ is the starred thickness to plasmon mean free path ratio and is equal to the intensity of the single plasmon component of the real EELS spectrum divided by the zero loss intensity of the spectrum, $\sigma_{k,q}\left( \Delta - \left( j-1\right) E_{p}, \beta\right)$ is calculable theoretically and is the partial inelastic scattering cross section relative to the inelastic scattering process $q$ of the $k$ chemical species calculated for the energy range $\Delta - \left( j-1\right) E_{p}$, $j$ = 1, 2, $\cdots$; this latter is the actual energy loss range over which the $j$th inelastic scattering intensity term $I_{k,q}^{*,j}$ of the characteristic edge $(k,q)$ is summed; $I_{k,q}^{*,j}$ is made up of electrons having been scattered inelastically $j$ times one of which is through the inelastic scattering process $(k,q)$ and $(j - 1)$ times are through the plasmon process; $E_{p}$ is the plasmon energy; and $\beta$ is the experimental collection semi-angle];
 \item 	\textbf{$X_{i}=t/\lambda_{i}$, \emph{the starred thickness to total inner-shell inelastic scattering mean free path ratio}}, from \cite{Hadji-2018a} one has:
\begin{equation}
X_{i}=\frac{t}{\lambda_{i}} = \sum_{k}\sum_{q}n_{k}^{*,a}\sigma_{k,q}^{\textrm{AS}} \left( = \sum_{k}n_{k}^{*,a}\sum_{q}\sigma_{k,q}^{\textrm{AS}}= \sum_{k}\sum_{q}\frac{t}{\lambda_{k,q}^{*}}\right)  \label{15Fifteen}
\end{equation}
\textbf{\{}where $t$ is the specimen thickness, $\lambda_{i}$ is the total inner-shell inelastic scattering mean free path; $k$ is label referring to chemical species and is summed over all chemical species present within the specimen under study, $q$ refers to the inelastic scattering process(es) relevant to atom species $k$ and is summed over all possible inelastic scattering processes of chemical species $k$, $\sigma_{k,q}^{\textrm{AS}}$ is the $(k,q)$-inner shell inelastic scattering cross section for the full energy loss interval over which the $(k,q)$-inner shell related inelastic scattering intensity is distributed and is calculable by taking the asymptotic value of $\sigma_{k,q}$ corresponding to large $\Delta$ (see \cite{Egerton-1996}); and $\lambda_{k,q}^{*}$ is the starred inelastic scattering mean free path associated with the $(k,q)$-inelastic scattering process, therefore this $\lambda_{k,q}^{*}$ is the $(k,q)$-related mean free path observed when none of the inelastic scattering processes associated with the specimen under study is switched off (made inactive), thus $\lambda_{k,q}^{*}$ is as it can be obtained using the intensity of the single $(k,q)$-related inner shell component within the real experimental EELS spectrum, see \cite{Hadji-2018a} for details\textbf{\}}; and
  \item $t/\lambda_{p}^{*}$, \textbf{\emph{the starred thickness to plasmon mean free path ratio}}, is equal to the ratio of the plasmon single intensity within the real EELS spectrum to the zero loss intensity within also the real experimental spectrum.  $t/\lambda_{p}^{*}$ is given by:
\begin{equation}
\frac{t}{\lambda_{p}^{*}} = \frac{t}{\lambda_{T}} -X_{i} \left(=\frac{t}{\lambda_{T}}-\frac{t}{\lambda_{i}}\right), \label{16Sixteen}
\end{equation}
\textbf{{\large [}}where $\lambda_{p}^{*}$ is the starred plasmon mean free path and $\lambda_{T}$ is the total inelastic mean free path; $t/\lambda_{T}\left( = t/\lambda_{T}(\beta) \right)$ is the thickness to total inelastic mean free path ratio and relates to the real EELS spectrum only\textbf{{\large ]}}.
 \end{itemize}
To extract the starred NAAs, $n_{k}^{*,a}$, and the starred thickness to plasmon mean free path ratio, $t/\lambda_{p}^{*} $, one, therefore, starts with making available the various parameters, $\sigma_{k,q}^{eff}\left( \Delta, \beta\right)$, $n_{k}^{a,NM}$, $\sigma_{k,q}^{AS}$ and $t/\lambda_{T}(\beta)$,  $\cdots$
\item[2)] \textbf{the correction for the unwanted effects of the combination of poly and plural inelastic scatterings} of these $n_{k}^{*,a}$ and $t/\lambda_{p}^{*} $ starred EELS parameters to get to their True counterparts, i.e. the $n_{k}^{a}$ and $t/\lambda_{p} $ free from the said unwanted effects. This can be achieved through two different, but fully equivalent, sets of formulas each of which relates to a different method, namely: Log Method and Ratio Method, see \cite{Hadji-2018a}. Here, use is made of the formulas associated with Log Method, i.e. the following formulas:
\begin{description}
\item[a)] \textbf{\emph{the plasmon process-related formula for correction for the unwanted effects}}:
\begin{equation}
\frac{t}{\lambda_{p}}=-\ln \left\lbrace \left[1+ \frac{X_{i}}{t/\lambda_{T}}\left( \exp \left( \frac{t}{\lambda_{T}}\right) -1\right)  \right]\exp \left( -\frac{t}{\lambda_{T}}\right)  \right\rbrace  \label{17Seventeen}
\end{equation}
which yields $t/\lambda_{p}$, the ``true'' \index{thickness(es)}thickness to plasmon \index{mean free path}mean free path ratio, through insertion of $X_{i}$, the difference between the thickness to total inelastic mean free path ratio, $t/\lambda_{T}$, and the starred thickness to plasmon mean free path ratio, $t/\lambda_{p}^{*}$, see (\ref{16Sixteen});  $X_{i}$ represents the sum of the thickness to inner-shell-related mean free path ratios $t/\lambda_{k,q}^{*}$, $\lambda_{k,q}^{*}$ being the starred inelastic scattering mean free path associated with the inner-shell inelastic scattering process ($k$,$q$) and the sum is over all inelastic scattering processes relevant to the atom species present within the volume of matter the spectrum under study is from;
\item[b)]  \textbf{\emph{the formula relative to the inelastic scattering inner-shell process $q$ of the atom species $k$}} (i.e. the $(k,q)$-related inner-shell process) \textbf{\emph{for correction for the unwanted effects}}:
\begin{equation}
\frac{t}{\lambda_{k,q}} = -\ln \left\lbrace \left[1+ \frac{X_{T}-X_{k,q}^{*}}{X_{T}}\left( \exp \left( X_{T}\right) -1\right)  \right]\exp \left( -X_{T}\right)  \right\rbrace=n_{k}^{a}\sigma_{k,q}^{AS}, \label{18Eighteen}  
\end{equation}
where ($k,q$) is the label of the inelastic scattering process $q$ of the chemical species $k$. Hence, the insertion into this (\ref{18Eighteen}) of values for $X_{T}=t/\lambda_{T}$ and $X_{k,q}^{*}=t/\lambda_{k,q}^{*}= n_{k}^{*,a}\sigma_{k,q}^{AS}$, both extracted from the same real EELS spectrum, gives the \index{thickness(es)}thickness to $(k,q)$-related inner-shell \index{mean free path}mean free path ratio $X_{k,q}=t/\lambda_{k,q}$ with no \index{poly}poly scattering-related effects. Thereafter, the division of the resulted $t/\lambda_{k,q}$ by $\sigma_{k,q}^{AS}$, the asymptotic value for the $(k,q)$-related inner-shell inelastic scattering cross section, yields $n_{k}^{a}$, the True number of $k$ atoms per unit area of specimen, since $t/\lambda_{k,q}=n_{k}^{a}\sigma_{k,q}^{AS}$; 
\end{description}
and, finally,
\item[3)] \textbf{the deduction, using (\ref{11Eleven}), of the various true PSCAs}, $\sigma_{k,q}(\beta)$, from the true EELS parameters obtained in \textbf{2)}, i.e. from the various (true) $n_{k}^{a}$ and $t/\lambda_{p}$.
\end{description}
Thereafter, if the material under study is not monatomic, using (\ref{05Five}) allows deducing the chemical content ratio(s) from the various true PSCAs obtained in \textbf{3)} and then forming, using(\ref{10Ten}), the non-linear essential equation to get, on solving numerically, the electron concentration from.
Thus, the determination of the electron concentration requires doing calculations which are much more easily achieved with the help of a computer, e.g. through the program\footnote{This program is now usable through eelsMicr, version 5.} related to \cite{Hadji-2016} or using the appropriate sub options of the fifth version of eelsMicr, see \cite{Hadji-2020a} for the code and the user's manual and \cite{Hadji-2020b} for the executable. 
\section{Application: an example using a silicon nitride spectrum}
\label{ApplPractEx}
\begin{figure}\begin{center}
\includegraphics[scale=.55]{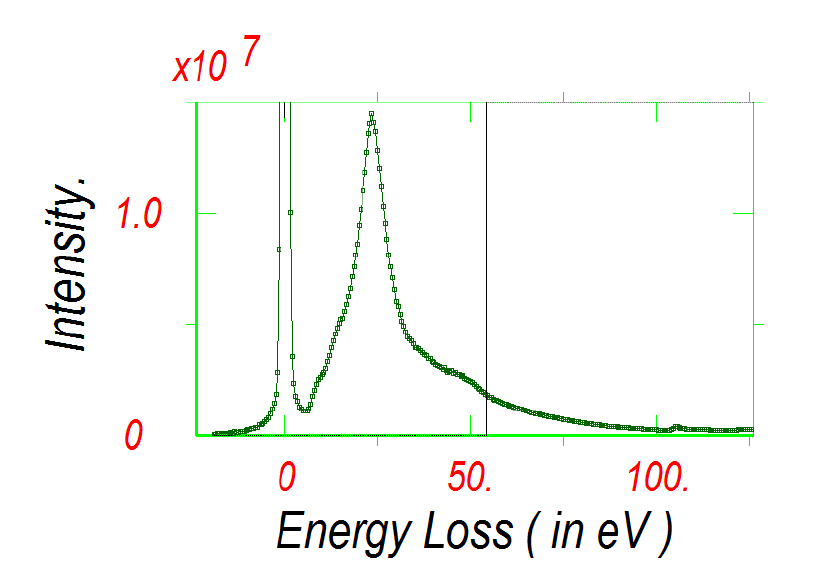}  
\caption{Portion of the merged EELS spectrum showing a vertical line, located at energy loss 54.5 eV, used to separate the low loss part scaled during the merging operation from the remaining non-scaled part.}
\label{Fig1}
\end{center}
\end{figure}
\subsection{Intensity determination from the silicon nitride spectrum considered and correction for incident beam convergence}
The data of Table \ref{tabl1Si3N4} were extracted from the silicon nitride EELS spectrum, a portion of which is shown in figure \ref{Fig1}, obtained by merging powder silicon nitride-related low loss and core loss spectral data by Srot, \cite{Srot-2008}. These were collected under a collection semi-angle of 6.5 mrad using an incident beam of 100 keV electrons with a convergence semi-angle of 10 mrad. The merging point, at $E_{M}=54.5$ eV, is indicated in the figure by a vertical line separating the spectrum scaled and non-scaled parts. The merging of the two spectra was done using a least squares fitting procedure in which the core loss spectrum and a constant function were fitted to the low loss spectrum over the energy interval ranging from 54.5 to 66 eV (24 data points were used). The fitting operation gave a fitting factor of $5.1882\times 10^{3}$ and a constant function of 440.2; and the merging was realized by subtracting the obtained constant function from the low loss spectrum, scaling the resulting spectrum by $S_{F} = 192.74$, the inverse of the fitting factor, and transferring to the core loss spectrum the data points of the thus resulting adjusted low loss spectrum from 54. eV and down to form the full electron energy loss spectrum. On the other hand, the fact that the full spectrum constructed in this way is made up of two parts of which one is a scaled portion of another spectrum has scaling-related consequences which, for a proper determination of the uncertainties on the various values relevant to the various low loss, total, edge -intensities extractable from this full spectrum, need be taken into account. Effectively, here the intensity of the part of the spectrum, the low loss part up to the energy loss $(E_{M} - 0.5)$ eV, having been subjected to scaling through multiplication by a factor, noted $S_{F}$ above, (a) the uncertainty on any fraction $I_{s}$ of this part is uncertain by an amount given by $\Delta I_{s}=S_{F} \sqrt{I_{s}/S_{F} }=\sqrt{S_{F} I_{s} }$ while (b) any portion $I_{u}$ of the part of the non-scaled spectrum is uncertain by the amount $\Delta I_{u}=\sqrt{I_{u}}$. Therefore, the uncertainty on an amount of intensity $I_{x} = I_{s} + I_{u}$ resulting from summing two fractions one is relative to the scaled part and the other one is relative to the non-scaled part is not given by $\sqrt{I_{x}}$ but is given by $\Delta I_{x}=\Delta I_{s}+\Delta I_{u}=\sqrt{S_{F} I_{s}}+\sqrt{I_{u}}$. This is the expression used in Option 0 of version 5 of eelsMicr, see \cite{Hadji-2020a}, and this Option 0 was employed to get the uncertainty-related figures (as well as the other figures in fact) of Table \ref{tabl1Si3N4}. 
\begin{table}[!htb]
\centering
 \textbf{\caption{\label{tabl1Si3N4} Intensity Data.}}
 \begin{minipage}{15.1cm}
 The Here Given Data were Obtained on Removing Background Under Nitrogen K-edge Without Previous Subtraction of Background Under Silicon L-edge. These Are Virtually the Same As the Data Obtained After Subtraction of Background Under Silicon L-edge. Usual $AE^{r}$ Power Law Model Used for Under Edge Background Estimate and Subtraction; Fitting Intervals Used: [80, 92.5] eV for Silicon L-edge and [299.5, 360] eV for Nitrogen K-edge. AIntensities Integrated Over Energy Window Width $\Delta$=100 eV: $I_{Si,L-ed}$ ($\Delta$) Is Silicon L-Edge Related Intensity, $I_{N,ed}$ ($\Delta$) Is Nitrogen K-Edge Related Intensity and  $I_{LowLoss}$ ($\Delta$) Is Low Loss Intensity.  $I_{ZeroLoss}$ Is Zero loss Intensity and  $I_{Total}$ Is Total Intensity Within EELS Spectrum;. (ed = edge), Incident Electron Energy = 100 keV. Specimen Name: VG HB501UX; Material: Si$_{3}$N$_{4}$; Source/Purity: powder, Submitted by Vesna Srot, January 21, 2008. Author Comments: Analyst: Lingyang Li. Temperature: Room. Convergence Semi-angle = 10 mrad and Collection Semi-angle = 6.5 mrad See \cite{Srot-2008}. The Calculations of Errors and Uncertainties Were Carried Out Taking Account of the Consequences of the Scaling Operation Made to Merge the Low Loss and the Core Loss Original Spectra. The Correction For Incident Beam Convergence, Line Three, Was Done Using An Adaptation of Program ConCor2, See \cite{Egerton-1996}, Installed in Option 0 of eelsMicr Version 5, See ,\cite{Hadji-2020a}, \cite{Hadji-2020b}.\\
 \rule[0mm]{15.1cm}{3.pt}\\
 \end{minipage} \\
 \begin{tabular}{cccccr}
CFBC &$I_{Si,L-ed}$     &  $I_{N,ed}$        &  $I_{ZL}=I_{ZeroLoss}$   & $I_{LL}=I_{LowLoss}$ & $I_{T}=I_{Total}$ \\ 
     &$\Delta=$100 eV   &  $\Delta=$100 eV  &        &  $\Delta=$100 eV     &     \\
&&&&\\      
\hline

&&&&\\ 
   & 2.549$\times 10^{7}$  &  1.055$\times 10^{6}$   &	 1.190$\times 10^{9}$  & 1.763$\times 10^{9}$ & 1.822$\times 10^{9}$   \\

No &  $\pm$  5.1$\times 10^{3}$  &  $\pm$  1.0$\times 10^{3}$ &  $\pm$4.8$\times 10^{6}$  &  $\pm$  5.8$\times 10^{6}$ & $\pm$ 5.3$\times 10^{6}$   \\
   &    (0.02 \%)    &   (0.1 \%)       &   (0.04 \%)      &  (0.03 \%)     &  (0.03 \%)  \\ 
%
&&&&\\
 \hline
&&&&\\ 
%
     &  4.977$\times 10^{7}$      &	 1.694$\times 10^{6}$   & 	2.817$\times 10^{9}$  & 4.173$\times 10^{9}$ & 4.313$\times 10^{9}$   \\
 Yes & $\pm$  9.9$\times 10^{3}$  &	 $\pm$  1.6$\times 10^{3}$    &   $\pm$1.1$\times 10^{6}$  &  $\pm$  1.4$\times 10^{6}$  & $\pm$ 1.38$\times 10^{4}$   \\
     &  (0.02 \%)  &   (0.1 \%)    &   (0.04 \%)         &  (0.03 \%)  &  0.03 \%)  \\ 
%
& &  &  &  \\
\hline  
\end{tabular}
\centering
\begin{tabular}{l}
\begin{minipage}{13.cm}
{\rule{0pt}{2ex}\small CFBC= Correction For Beam Convergence, No = correction for beam convergence not done, Yes = correction for beam convergence done.}
\end{minipage}
\end{tabular}
\end{table}
In this Table \ref{tabl1Si3N4}, $I_{Si,L-ed} (\Delta)$, $I_{N,ed} (\Delta)$ and $I_{LowLoss} (\Delta)$ are, respectively, the silicon L-edge-related, the nitrogen K-edge-related and the low loss intensities, each summed over energy window width $\Delta = 100$ eV of the merged spectrum.  $I_{ZeroLoss}$ is zero loss intensity summed over energy interval up to 5 eV of the merged spectrum and $I_{Total}$ is the total intensity within the merged EELS spectrum. The lines ``No'' and ``Yes''  of this Table \ref{tabl1Si3N4} compare  data, respectively, not-corrected and corrected for incident beam convergence effects. The data of line ``Yes'' were obtained from the data of line ``No'' using an adaptation of the program ConCor2 from \cite{Egerton-1996} included into the code of Option ``O'' of eelsMicr version 5. The intensity values of line Yes of this Table \ref{tabl1Si3N4} are the first key EELS data extracted from the EELS spectrum under analysis needed for producing the next suitable EELS parameters, given in line Yes of Table \ref{tabl2Si3N4}, which are required for the determination of the next intermediate proper EELS parameters, given in line SAM of Table \ref{tabl3Si3N4}; that is the parameters required for the production of the data strictly necessary for obtaining the true EELS parameters, presented in line 'True' of Table \ref{tabl3Si3N4}, which are essential for getting to the characterizing physical parameters, attainable through different analytical methods, associated with the material the EELS spectrum considered is from. 
\begin{table}[!htb]
\centering
 \textbf{\caption{\label{tabl2Si3N4} Data Needed to Apply SAM.}}
 \begin{minipage}{15.1cm}
 Data Calculated Using the Measured Intensity Values of Table 1: $t/\lambda_{T}$ Is the Thickness to the Total Inelastic Mean Free Path Ratio , $(t/\lambda_{p})^{\textrm{NM}}$ Is the Thickness to Plasmon Mean Free Path Ratio According to the Normal EELS Microanalysis Method,  $I_{Si,L-ed}/I_{LL}$ Is the Silicon L-edge Intensity Measured Over a 100 eV Energy Window Width Starting from the Edge Threshold Energy to the Low Loss Intensity, Integrated Over the Same Energy Window width, Ratio and $I_{N,ed}/I_{LL}$ Is the Nitrogen K-edge Intensity Measured Over, Also, a 100 eV Energy Window Width Starting from the Edge Threshold Energy to the Low Loss Intensity, Integrated Over the Same Energy Window Width, Ratio. The Correction For Incident Beam Convergence, Line Three, Was Also Done Using an Adaptation of Program ConCor2, See \cite{Egerton-1996}, Used in Option 0 of eelsMicr Version 5, See \cite{Hadji-2020a}, \cite{Hadji-2020b}. Data Corrected for Incident Beam Convergence using $E_{k}+ \Delta/2$.\\
 \rule[0mm]{15.1cm}{3.pt}\\
 \end{minipage} \\
 \begin{tabular}{lcccr}
CFBC  &  {\Large $\frac{t}{\lambda_{T}}$} &  {\Large $(\frac{t}{\lambda_{p}})^{\textrm{NM}}$}      &   {\Large $\frac{I_{Si,L-ed}}{I_{LL}}$}         & {\Large $\frac{I_{N-ed}}{I_{LL}}$} \\ 

&&  & &    \\
 \hline
&&&\\
%
%
     &  0.426             &	0.393                &	1.445$\times 10^{-2}$        & 5.985$\times 10^{-4}$   \\
 No  & $\pm$  8.27$\times 10^{-5}$  &	 $\pm$  7.38$\times 10^{-5}$    &  $\pm$  7.61$\times 10^{-6}$   &  $\pm$  7.79$\times 10^{-7}$   \\
      & (0.02 \%)         &   (0.02 \%)       &  (0.05 \%)           &  (0.1 \%)  \\ 
  \hline
 & &  &  &  \\
%
%
      &    0.426          &    0.393           &	1.192$\times 10^{-2}$        & 4.058$\times 10^{-4}$   \\
 Yes  & $\pm$  8.27$\times 10^{-5}$  &	 $\pm$  7.38$\times 10^{-5}$    &  $\pm$  6.28$\times 10^{-6}$   &  $\pm$  5.28$\times 10^{-7}$   \\
      &  (0.02 \%)        &    (0.02 \%)       &  (0.05 \%)             &  (0.1 \%)  \\ 
%
 & &  &  &  \\
\hline   
\end{tabular}
\end{table}
\subsection{Determination of the data needed to apply SAM}
The lines 'No' and 'Yes' of Table \ref{tabl2Si3N4} compare data, respectively, non-corrected and corrected for the effects of incident beam convergence obtained, respectively, from the data of lines 'No' and 'Yes' of Table \ref{tabl1Si3N4} through, also, Option 0 of version 5 of eelsMicr:
\begin{description}
\item[$\circ$] 	$t/\lambda_{T}  [=\ln(I_{T}/I_{ZL} ) ]$ is the thickness, $t$, to the total inelastic mean free path, $\lambda_{T}$, ratio obtained through the so called log-ratio model and is a parameter greatly independent of incident beam convergence $\alpha$; this is for the reason that it comes from the ratio of two intensities, namely $I_{T} (\beta)$ and $I_{ZL} (\beta)$ for parallel illumination collected using collection semi-angle $\beta$, which, according to the model by \cite{Craven-et-al-1981} and \cite{Scheinfein-Isaacson-1984} can be related to $I_{T} (\alpha,\beta)$ and $I_{ZL} (\alpha,\beta)$, respectively, the total and zero loss intensities from a convergent incident beam with convergence semi-angle $\alpha$ collected using also collection semi-angle $\beta$ through $I_{T} (\beta)\approx(\alpha/\beta)^{2} I_{T} (\alpha,\beta)$ and $I_{ZL} (\beta)\approx(\alpha/\beta)^{2} I_{ZL} (\alpha,\beta)$, i.e. like, see for instance \cite{Egerton-1996} for details, the low loss intensities $I_{LL} (\beta)$ and $I_{LL} (\alpha,\beta)$ are, thus giving a $t/\lambda_{T} =\ln \left[ I_{T} (\beta)/I_{ZL} (\beta)\right]  \approx \ln \left[I_{T} (\alpha,\beta)/I_{ZL} (\alpha,\beta)\right]$ approximately independent of $\alpha$, which is much helpful as regards to extracting corrected for incident beam convergence starred EELS quantities from a real EELS spectrum obtained using a $\alpha/\beta<2$, i.e. quantities necessitating corrections; in effect, this allows using (\ref{16Sixteen}) together with SAM to determine the starred thickness to plasmon mean free path ratio, $t/\lambda_{p}^{*}$, corrected for incident beam convergence though injection of (i) the value for  $t/\lambda_{T}$  obtainable using the log-ratio model together with the zero loss and the total intensities as measured directly from the experimental EELS spectrum and (ii) the  $t/\lambda_{i}$  value corrected for the effects of $\alpha$ using the relevant $F_{1}$ correction factors. These factors are usable together with inner-shell related inelastically scattered intensities and are connected with the model by \cite{Craven-et-al-1981} and \cite{Scheinfein-Isaacson-1984};
\item[$\circ$] 	$\left( t/\lambda_{p}\right) ^{NM} [=\ln(I_{LL}/I_{ZL} ) ]$ is the thickness, $t$, to plasmon mean free path, $\lambda_{p}$, ratio according to NEELSM; the experimental value for this parameter is not strictly necessary for the proper application of SAM but is a good first estimate for the value of the $t/\lambda_{p}^{*}$ looked for to use when applying SAM;
\item[$\circ$] $I_{Si,ed} (\beta,\Delta)/I_{LL} (\beta,\Delta)$ and $I_{N,ed} (\beta,\Delta)/I_{LL} (\beta,\Delta)$ are the ratios of the intensities of, respectively, the silicon L-edge and nitrogen K-edge measured over a same energy loss interval width $\Delta$ = 100 eV, starting at the observed chemically upward shifted L-edge threshold energy $\sim $ 103. eV for $I_{Si,ed} (\beta,\Delta)$ and at the observed chemically downward shifted nitrogen K-edge threshold energy $\sim $ 398 eV for $I_{N,ed} (\beta,\Delta)$, to the low loss intensity summed also over the same energy loss interval $\Delta$; the low loss intensity, $I_{LL} (\beta,\Delta)$, also free from the effects of the incident beam convergence need be obtained from the corresponding intensities not free from the incident beam convergence effects; in practice, the ratios free from these effects are proportional to the appropriate corresponding ratios involving non-corrected for incident beam convergence $I_{Si,ed} (\alpha,\beta,\Delta)$, $I_{N,ed} (\alpha,\beta,\Delta)$, and $I_{LL} (\alpha,\beta,\Delta)$ through the reciprocal of the suitable factor $F_{2} = (\alpha/\beta)^{2}F_{1}$ when $\alpha\geq\beta$ and $F_{2} = F_{1}$ when $\alpha\leq\beta$ with $F_{1} (=I_{Si,ed} (\alpha,\beta,\Delta)/I_{Si,ed} (\beta,\Delta),I_{K,ed} (\alpha,\beta,\Delta)/I_{K,ed} (\beta,\Delta)) <1$ is calculable, see \cite{Egerton-1996}, and represents the factor by which incident beam convergence reduces the core loss intensity collected under collection semi-angle $\beta$; the calculation of $F_{1}$ is done here using the CONCOR2 Fortran program form \cite{Egerton-1996}, this program uses the expression, see \cite{Egerton-1996}, by \cite{Scheinfein-Isaacson-1984};
\end{description}
Of these four parameters only the three $t/\lambda_{T}$, $I_{Si,ed} (\beta,\Delta)/I_{LL} (\beta,\Delta)$ and $I_{N,ed} (\beta,\Delta)/I_{LL} (\beta,\Delta)$ are strictly needed for the determination of the starred, and therefore of the true, parameters of Table \ref{tabl3Si3N4}; the value for $(t/\lambda_{p} )^{NM}$ of line Yes being a good, but not strictly necessary, first approximate estimate for the starred value of $t/\lambda_{p}$  to use together with SAM in the process of determining the various starred EELS parameters. 
\subsection{Comparing EELS data obtained through NEELSM, SAM and IEELSM }
Table \ref{tabl3Si3N4} compares data obtained through NEELSM, SAM and IEELSM.
The data of the three data lines of this Table \ref{tabl3Si3N4} were obtained using: 
\begin{description}
\item[$\centerdot$] NEELSM for line 2 (labeled NEELSM); in this method the NAAs, i.e. $n_{Si}^{a}$ for silicon and $n_{N}^{a}$ for nitrogen, were obtained through (\ref{13Thirteen}) and the PSCAs per atom species, i.e. $\sigma_{p}^{Si}$ and $\sigma_{p}^{N}$, respectively, for silicon and nitrogen, were obtained using (\ref{11Eleven}) together with the appropriate data of line 2 (i.e. those of columns three, four and five);
\item[$\centerdot$] SAM for line 3 (labeled SAM); in this method the thickness to plasmon mean free path ratio and the NAAs are starred ones and were obtained using SAM, see above and \cite{Hadji-2018a}, and the PSCAs per atom species of this line are starred PSCAs, $\sigma_{p}^{*,k}$, and were obtained through (\ref{11Eleven}) written as $\sigma_{p}^{*,k} (\beta)=t/\lambda_{p}^{*} (\beta)/n_{k}^{*,a} $ where $t/\lambda_{p}^{*} (\beta)$ and $n_{p}^{*,k}$ are, respectively, the starred thickness to plasmon mean free path ratio and NAA of this SAM line; and
\item[$\centerdot$] IEELSM for line 4 (labeled True); this line contains (a) the true values for the thickness to plasmon mean free path ratio, column three, and the NAAs, columns four and five, obtained through, respectively, (\ref{17Seventeen}) and (\ref{18Eighteen}) in which the values for the appropriate starred parameters of line SAM were inserted; (b) the nitrogen to silicon content ratio, $N/Si$, obtained through (\ref{05Five}) plus the values of the appropriate parameters of this line True; and, finally, (c) the PSCAs per atom of silicon and of nitrogen species given in columns, respectively, seven and eight, obtained through (\ref{11Eleven}) together with the true, so given in this line True, values for $t/\lambda_{p}$ (column three) and $n_{k}^{a}$,(columns four and five). 
The  values of the parameters relative to the three last columns of this line True are values accurately represented by the formulas (\ref{05Five}), for the ratio $N/Si$, and (\ref{07Seven}) - and therefore (\ref{10Ten}) - for $\sigma_{p}^{Si}$ and $\sigma_{p}^{N}$. This is for the reason that the values for these $\sigma_{p}^{Si}$ and $\sigma_{p}^{N}$ come from the values of the parameters of columns three, four and five of line True, and therefore from the right values for $t/\lambda_{p} (\beta)$ and $n_{k}^{a}$, see above, i.e. from those which would be obtained if the inelastic scattering were through just the plasmon inelastic scattering process for $t/\lambda_{p} (\beta)$ and (if it were through) an inner-shell related inelastic scattering process relevant to the $k$ atom species only for $n_{k}^{a}$, with $k$ = Si, N. Consequently, any of $\sigma_{p}^{Si}$ and $\sigma_{p}^{N}$ of line True can be inserted 
\begin{description}
\item[($\alpha$)] into (\ref{07Seven}) to form an equation analytically solvable in $n_{k}$, the absolute atom concentration of the $k$ atom species, if values for $\beta$ and $\theta_{E}$ are known, thereafter $n_{e}$ deduces through (\ref{03Three}) (this is the method used to get $n_{e}$ in \cite{Hadji-2018a} and corresponds to options 1 and 2 in eelsMicr version 4, see \cite{Hadji-2018b, Hadji-2018c}; in eelsMicr version 5 this method corresponds to the branches ``(B: Beta is less than ThetaC:  Yes, C: Use Technique 1 (analytical solution))?:    Yes'' of options 1 and 2, see \cite{Hadji-2020a, Hadji-2020b} and 
\item[($\beta$)] into (\ref{10Ten}) together with the content ratio $n_{q}/n_{k}$, $q(\neq k)$ = Si, N, to form a non-linear equation in $n_{e}$ which is numerically solvable if a value for $m$ is known, this is the method considered here, this method corresponds to branches ``(B: Beta is less than ThetaC:  Yes, C:  Use Technique 1 (numerical solution)?:   Yes)'' of options 1 and 2 of eelsMicr version 5, see \cite{Hadji-2020a, Hadji-2020b}; the data of Table \ref{tabl4Si3N4} and Table \ref{tabl5Si3N4} were obtained using Option 1 → (B: Beta is less than ThetaC:  Yes, C:  Use Technique 1 (numerical solution)?:   Yes) of eelsMicr version 5.
\end{description}
\end{description}
\begin{table}[!htb]
\centering
 \textbf{\caption{\label{tabl3Si3N4} Some EELS Parameters Obtained Using NEELSM, SAM and IEELSM.}}
 \begin{minipage}{\linewidth}
Some EELS Parameters Obtained by Means of Three Methods: NEELSM, SAM With No Correction for Poly and Plural Scatterings and IEELSM  = SAM With Corrections. The SAM Used the Single Scattering Intensities from the Experimental $Si_{3}N_{4}$ Spectrum of \cite{Srot-2008}. Incident Electron Energy 100 keV; Integration Window, $\Delta$, Was 100 eV. The Statistical Errors Are Small, So the Errors on the Various Parameters Given Here Are of Purely Theoretical Nature and Come from the Calculated Partial Cross-Sections Used: an Arbitrary $\sim$ 3 \% Error on All Calculated Inelastic Scattering Cross Sections Was Used. Incident Beam Convergence = 10 mrad, Collection Semi-Angle = 6.5 mrad. Energy Increment (EINC) = 20.\\
 \rule[0mm]{\linewidth}{3.pt}\\
 \end{minipage} \\
 \begin{tabular}{lccccccr}
          Method & {\Large $\frac{t}{\lambda_{T}}$} & {\Large $\frac{t}{\lambda_{p}}$} & {\large $n_{\textrm Si}^{a}$} &  {\large $n_{\textrm N}^{a}$} &     {\large $\frac{\textrm N}{\textrm Si}$} & {\large $\sigma_{p}^{Si}$}  & {\large $\sigma_{p}^{N}$} \\ 
&  &  & $ (10^{21} at/m^{2})$  & $ (10^{21} at/m^{2})$  &   & $ (10^{-22}$ m$^{2})$ & $ (10^{-22}$ m$^{2})$ \\
 &  &  & &  &  &  &\\
 \hline
&&&&&\\
%
 NEELSM    & 0.426         & 0.393 	       & 1.457        & 1.880        & 1.290        & 2.698  & 2.091 \\ 
           & $\pm$ 0.2 \% & $\pm$ 0.2 \% & $\pm$ 0.05 \% & $\pm$ 0.1 \% & $\pm$ 0.2 \% & $\pm$ 0.2 \% &  $\pm$ 0.3 \% \\  
            &  &  & &  &  &  &\\
\hline
            &&&&&\\

  SAM  & 0.426 	      & 0.408         &  1.567      & 2.102      & 1.342      & 2.604      & 1.941 \\
      & $\pm$ 0.2 \% & $\pm$ 0.2 \% & $\pm$ 3 \%  & $\pm$ 2 \% & $\pm$ 6 \% & $\pm$ 3 \% &  $\pm$ 2 \% \\
 
 &  &  & &  &  &  &\\
\hline
&&&&&\\
 
True    & None &  0.404 	    &  1.285       &  1.712      &  1.333      &  3.143     &  2.358  \\
        &  &  $\pm$ 0.3 \%	&  $\pm$ 3 \%   &  $\pm$ 2 \%  &  $\pm$ 5 \%  &  $\pm$ 3 \% &  $\pm$ 2 \% \\  
%
 &  &  & &  &  &  &\\
\hline    
\end{tabular}
\centering
\begin{tabular}{l}
\begin{minipage}{15.3cm}
{\rule{0pt}{2ex}\small NEELSM = normal EELS microanalysis method; SAM = Successive Approximations Method; IEELSM = improved EELS microanalysis method = SAM plus corrections for the unwanted effects of the combination of polyscattering and plural scattering.}
\end{minipage}
\end{tabular}
\end{table}
\subsection{Extraction of physical parameters characterizing the material the EELS spectrum considered was obtained from}
The values for the content ratio and PSCAs of line True of Table \ref{tabl3Si3N4} were used together with (\ref{10Ten}) and the value for the electron effective mass, $m$, given in column one of Table \ref{tabl4Si3N4} to get the remaining various figures of this Table \ref{tabl4Si3N4} and the data of Table \ref{tabl5Si3N4}. The value $8.771\times10^{-31}$ kg  used for $m$ is that which can be obtained through the branch ``Option 1 $\rightarrow$ (B:   Beta is less than ThetaC:   Yes, C:  Use Technique 1 (analytical solution) ?:     Yes)'' of eelsMicr version 5, so through the ``analytical solution'' method. This value was obtained making use of the True values for the PSCAs of Table \ref{tabl3Si3N4} together with (\ref{07Seven}) and the value for the plasmon energy 23.5 $\pm$ 1 ($\sim$ 4 \%) eV (measured from the considered merged EELS spectrum and used to calculate $\theta_{E}$) to obtain $n_{\textrm{Si}}$ and $n_{\textrm{N}}$ and use them to get $n_{e}$ through (\ref{03Three}) and then deduce $m$ through (\ref{02Two}). On the other hand, the value $9.90\times10^{-32}$ ($\sim$11 \%) kg for the uncertainty on $m$ is the value which yields a value for the uncertainty on the plasmon energy equal to 1.00 eV determined  through the branch ``Option 1 $\rightarrow $ (B:  Beta is less than ThetaC:  Yes, C: Technique 1 (numerical solution)?:  Yes)'', so through the ``numerical solution'' method, of eelsMicr version 5 and therefore relates to the numerical method described here. Except for the uncertainty values, the data of both Table \ref{tabl4Si3N4} and Table \ref{tabl5Si3N4} can be reached through either of the ``analytical solution'' and ``numerical solution'' methods.\\
\textbf{\textit{Material's chemical formula:}} The material's chemical formula deduces from the value for the content ratio of line True of Table \ref{tabl3Si3N4}. This formula can be written as A$_{x}$B$_{y}$ with A=Si, B=N, $x=3$ and $y=3.998(=4.00)$ i.e. as A$_{3}$B$_{3.998(=4)}$, these values were obtained through Option 1 of eelsMicr version 5.\\
\textbf{\textit{Material's molar and molecular masses:}} The values for $x$ and $y$ relative to this formula lead to a molar mass $M$ = 139.97 g/mole, obtained using $M=xm_{x}+ym_{y}$ where $m_{x}$=28 g/mole and $m_{y}$=14 g/mole are the molar masses of the chemical elements, respectively, A = Si and B = N and to molecular mass of $M_{\textrm{m}} = M/N = 2.3232\times10^{-22}$ g, $N$ being the Avogadro number $6.025\times10^{23}$.\\
\textbf{\textit{Other physical parameters: These are given in Table \ref{tabl4Si3N4} and Table \ref{tabl5Si3N4}.}}\\
In Table  \ref{tabl4Si3N4}, 
\begin{description}
\item[a)] $n_{\textrm{Si}}$ and $n_{\textrm{N}}$ are the absolute atom concentrations of, respectively, silicon and nitrogen and were obtained using (\ref{09Nine}) together with the value for $n_{e}$, the electron concentration, given in Table \ref{tabl4Si3N4}; each of the values $3.613\times 10^{+28}$ at/m$^{3}$ and $4.816\times 10^{+28}$ at/m$^{3}$ of line IEELSM corresponding, respectively, to $n_{\textrm{Si}}$ and $n_{\textrm{N}}$ given in Table \ref{tabl4Si3N4} leads to (j) a number of molecules Si$_{3}$N$_{3.998(=4)}$ per unit volume, $n_{\textrm{M}}$, given by, e.g. see \cite{Hadji-2002}, $n_{\textrm{M}}=n_{\textrm{Si}}/x=n_{\textrm{N}}/y = 1.2044\times10^{22}$ molecules/cm$^{3}$ and (jj) a density for the material under study equal to $\rho = n_{\textrm{M}} M_{\textrm{m}} = 2.80 \pm 0.15$ g/cm$^{3}$; this density value is the same as the value, 2.80 $\pm $ 0.3 g/cm$^{3}$, for the ``radio frequency power 120 W, gas pressure 600 mTorr and 300 °C temperature''  material by \cite{Huang-et-al-2006} and compares well with the value $\sim 2.71\pm0.11$ g/cm$^{3}$ by \cite{Huszanka-et-al-2016} obtained for low pressure chemical vapor deposited Si$_{3}$N$_{4}$ (silicon nitride), the largest difference between any pair of the three of them being $\sim $ 3 \% only; the three values were obtained through different experimental means; the value 2.80$\pm$0.3 g/cm$^{3}$ was obtained using a microbalance system and the $\sim $ 2.71 $\pm $  0.11 g/cm$^{3}$ value was determined and confirmed, see \cite{Huszanka-et-al-2016}, through experimental techniques other than EELS, i.e. determined by Rutherford backscattering spectroscopy (RBS) and profilometry technique and verified by scanning transmission ion microscopy (STIM); on another hand, the density result for Si$_{3}$N$_{4}$ obtained here, i.e. using the model by \cite{Craven-et-al-1981} and \cite{Scheinfein-Isaacson-1984} for dealing with the effects of the incident beam convergence, is satisfactory to the extent that one can consider the model by \cite{Craven-et-al-1981} and \cite{Scheinfein-Isaacson-1984} is suitable for use together with SAM and, therefore, with IEELSM; 
\item[b)] 	the plasmon mean free path, $\lambda_{p}$, can be obtained through either of (\ref{01One}) and (\ref{08Eight}); 
the specimen thickness, $t$, can be obtained either using $t=[(t/\lambda_{p})_{exp} ] (\lambda_{p} )_{exp}$ where $(t/\lambda_{p} )_{exp}=0.4038$, see Table \ref{tabl3Si3N4}, and $(\lambda_{p} )_{exp}$=88.1, see Table \ref{tabl4Si3N4}, or using $t=(n_{k}^{a} )_{exp}/(n_{k} )_{exp}$ where, see line True of Table \ref{tabl3Si3N4}, $(n_{k}^{a} )_{exp}= 1.285\times10^{21}$ (or $=1.712\times10^{21}$ ) at/m$^{2}$ and $(n_{k} )_{exp}= 3.613\times10^{28}$ (or = $4.816\times10^{21}$  ) at/m$^{3}$;
\item[c)] the value for the plasmon energy, $E_{p}$, was obtained through (\ref{02Two}) using the data for $m$ and $n_{e}$ of this Table \ref{tabl4Si3N4}.
\end{description}
\begin{table}[!htb]
\centering
 \textbf{\caption{\label{tabl4Si3N4} Some True Physical Parameters for Silicon Nitride.}}
 \begin{minipage}{\linewidth}
Part of the True Experimental Parameters Obtained from Merged Spectrum By Srot, \cite{Srot-2008}, for Si$_{3}$N$_{4}$ Through IEELSM. The Errors on these Results Are for the Most Part of Theoretical Nature. Following Branch ``Option 1 $\longrightarrow$ Beta less than ThetaC $\longrightarrow$ Technique 1 (numerical solution)'' of eelsMicr Version 5 was Used to Get the Data of This Table.\\
 \rule[0mm]{\linewidth}{3.pt}\\
 \end{minipage} \\
 \begin{tabular}{ccccccr}
 $m$ & {\large $n_{e}$} & {\large $n_{\textrm Si}$}      &  {\large $n_{\textrm N}$}      & {\large $\lambda_{p}$}  & {\large $t$} &  {\large E$_{p}$} \\ 
 $(10^{-31})$kg & ($10^{29}$ m$^{-3}$)&  $ (10^{28}$ at/m$^{-3}$)& $ (10^{28}$ at/m$^{-3})$   & (nm)  & (nm) & (eV)   \\
  &&  & &  &  &  \\
 \hline
 &&&&&&\\
%
 
 8.771 & 3.853 &  3.613       &  4.816      &    88.1  &  35.6  & 23.5 	    \\
  $\pm$ 11 \%& $\pm$ 3 \% &  $\pm$ 3 \%   &  $\pm$ 3 \%   &  $\pm$ 3 \%  & $\pm$ 0.8 \%  & $\pm$ 1. \\  
%
  &&  &  & &  & \\
\hline    
\end{tabular}
\end{table}
Table \ref{tabl5Si3N4} gathers additional results of which some relate to the so called random phase approximation (RPA), these can be obtained through both of the collection semi-angle-less-than-ThetaC-related analytical and numerical methods of option 1 available in eelsMicr version 5, and some others relate to the exchange and correlation (ExCor) and are reachable through the analytical method only.\\
In this Table \ref{tabl5Si3N4},
\begin{description}
\item[i)] 	$k_{F}$, given by $k_{F}=(3\pi^{2} n_{e} )^{1/3}$, is the Fermi vector, $E_{F}$,  given by (\ref{02Two}), is the Fermi energy and $\theta_{E}$, given by (\ref{06Six}), is the characteristic angle for energy $E$, are independent of the case of RPA and ExCor considered, but
\item[ii)] all of $\alpha$, the plasmon dispersion coefficient, $k_{c}$, the plasmon cutoff vector and $\theta_{c}$, the plasmon cutoff angle, depend on the case, RPA or ExCor, considered. The expressions for $\alpha$, see \cite{Egerton-1996}, are: in the RPA case
\begin{equation}
\alpha_{RPA}=\frac{3E_{F}}{5E_{p}}, \label{19Nineteen}
\end{equation}
and in the ExCor case
\begin{equation}
\alpha_{ExCor}=\frac{3E_{F}}{5E_{p}} \left[ 1-\left( \frac{E_{p}}{4E_{F} }\right) ^{2} \right] . \label{20Twenty}  
\end{equation}
\noindent The expression for $k_{c}$, see \cite{Hadji-2016}, is:
\begin{numcases}{  k_{c}=}
\frac{k_{F}}{2\alpha-1} \left( 1-\sqrt{1-\frac{\left( 2\alpha-1\right) E_{p}}{E_{F}}} \right)   &  for $\alpha\neq 1/2, $ \nonumber \\ 
 \frac{m_{0} E_{p}}{\hbar^{2} k_{F}}=\frac{k_{F} E_{p}}{2E_{F}} & for $\alpha=1/2. $ \label{21TwentyOne}
\end{numcases} 
\end{description}
where $\alpha$ is given either by (\ref{19Nineteen}) or by (\ref{20Twenty}) according to which of the RPA and ExCor cases is considered. $\theta_{c}$, is as follows:
\begin{equation}
\theta_{c}\simeq\frac{\sqrt{k_{c}^{2}-\theta_{E}^{2} k_{0}^{2}} }{k_{0}^{2}},  \label{22TwentyTwo}                                                       
\end{equation}
where $k_{c}$ is given by (\ref{21TwentyOne}), $\theta_{E}$ is given by (\ref{06Six}) and $k_{0}$ is the wave vector of the incident electron beam.
\begin{table}[!htb]
\centering
 \textbf{\caption{\label{tabl5Si3N4}More True Results for Si$_{3}$N$_{4}$}}
 \begin{minipage}{\linewidth}
Extra True Experimental Results Obtained for Si$_{3}$N$_{4}$ in RPA and ExCor. The Errors on these Results Are for the Most Part of Theoretical Nature. Following Branch ``Option 1 $\longrightarrow$ Beta less than ThetaC $\longrightarrow$ Technique 1 (numerical solution)'' of eelsMicr Version 5 was Used to Get the Data of This Table.\\
 \rule[0mm]{\linewidth}{3.pt}\\
 \end{minipage} \\
 \begin{tabular}{lccccccl}
 Method    &   {\large $k_{\textrm F}$}      & \textrm{ \quad   } {\large $E_{\textrm F}$}        & \textrm{ \quad     \quad } {\large $\theta_{E}$}          &  \textrm{ \quad    }   {\large $\alpha$}   & \textrm{ \quad     \quad }{\large $k_{c}$}     & \textrm{ \quad   }{\large $\theta_{c}$} \\ 
           &   (10$^{10}$ m$^{-1}$)              & \textrm{ \quad   }(eV)                   & \textrm{ \quad     \quad } (mrad)     &\textrm{ \quad        \quad }& \textrm{ \quad    }(10$^{10}$ m$^{-1}$) & \textrm{ \quad     \quad } (mrad) \\
 &  &  & &  &   &\\
 \hline
&&&&&\\
%
 RPA  & 	2.251   & \textrm{ \quad     \quad }	19.32    & \textrm{ \quad     \quad }    0.128   &  \textrm{ \quad     \quad }   0.493    &    \textrm{ \quad     \quad }  1.363  &      \textrm{ \quad   } 8.03  \\
	  &  $\pm$ 0.9 \% & \textrm{ \quad     \quad } $\pm$ 2 \% & \textrm{ \quad     \quad } $\pm$ 4 \% & \textrm{ \quad     \quad } $\pm$ 6 \%  & \textrm{ \quad    \quad } $\pm$ 3 \% &  \textrm{ \quad   }$\pm$ 3 \%  \\  
 &&&&&&\\
\hline
 &&&&&&\\
%
ExCor  &	    2.251   & \textrm{ \quad     \quad }   19.32    &   \textrm{ \quad     \quad }   0.128  &    \textrm{ \quad     \quad }   0.448  &   \textrm{ \quad    \quad }    1.328   &    \textrm{ \quad    }     7.82  \\
       &  $\pm$  0.9 \% & \textrm{ \quad     \quad } $\pm$ 2 \% & \textrm{ \quad     \quad } $\pm$ 4 \% &   \textrm{ \quad     \quad }  $\pm$ 7 \% &	\textrm{ \quad     \quad } $\pm$ 3 \% &	\textrm{ \quad    } $\pm$ 3 \%  \\
 &&&&&&\\
\hline  
\end{tabular}
\end{table}
\textbf{\textit{Additional result extractions:}} Many physical quantities that are free from any unwanted effects associated with the combination of poly and plural inelastic scatterings and that are all entirely relevant to one another have been extracted from a single EELS spectrum. However, one can go on and produce from the same spectrum others properties fully relevant to the ones already produced. These properties include optical properties of the material the spectrum considered is from, and a way of getting them can be that described and, then, applied to a spectrum from a boron nitride specimen in \cite{Hadji-2018d}.

\section{Conclusion}
The extraction of a number of EELS and physical parameters, free from any unwanted effects associated with the combination of poly and plural inelastic scatterings, using a technique different, numerical, from that, analytical, employed in \cite{Hadji-2018a} has been reported in minute detail. Also, here the extraction is from a different kind of EELS data, i.e. data \textit{requiring corrections for incident beam convergence}. The silicon nitride data by \cite{Srot-2008} was used. The obtained density-related result of $2.80\pm 0.15$ g/cm$^{3}$ compares very well indeed with existing corresponding results. The software used, eelsMicr version 5, is an extended version of eelsMicr version 4 and is easily reachable.\\
One can also say that the experimental results obtained here for silicon nitride confirm the validation of the improved electron energy loss spectroscopy microanalysis method, \cite{Hadji-2018a}, reached through other experimental results from boron nitride and amorphous hydrogenated silicon.
\section*{Acknowledgments}
Nothing to declare.

%
\end{document}